\documentclass[pra,aps,preprint,showpacs]{revtex4}
\usepackage{bm}
\usepackage{amsmath,amssymb}
\usepackage{graphicx}

\begin{document}

\title{Siegert pseudostates: completeness and time evolution}
\author{Robin Santra}
\author{Jeffrey M. Shainline}
\author{Chris H. Greene}
\affiliation{Department of Physics and JILA, University of Colorado,
Boulder, CO 80309-0440, USA}
\date{\today}
\begin{abstract}
Within the theory of Siegert pseudostates, it is possible to accurately calculate bound states 
and resonances. The energy continuum is replaced by a discrete set of states. Many questions of
interest in scattering theory can be addressed within the framework of this formalism, thereby
avoiding the need to treat the energy continuum. For practical calculations it is important to 
know whether a certain subset of Siegert pseudostates comprises a basis. This is a nontrivial 
issue, because of the unusual orthogonality and overcompleteness properties of Siegert pseudostates.
Using analytical and numerical arguments, it is shown that the subset of bound states and outgoing
Siegert pseudostates forms a basis. Time evolution in the context of Siegert pseudostates is also
investigated. From the Mittag-Leffler expansion of the outgoing-wave Green's function, the 
time-dependent expansion of a wave packet in terms of Siegert pseudostates is derived. In this
expression, all Siegert pseudostates---bound, antibound, outgoing, and incoming---are
employed. Each of these evolves in time in a nonexponential fashion. Numerical tests underline
the accuracy of the method.
\end{abstract}
\maketitle

\section{Introduction}
\label{sec1}

Let us consider the following model potential:
\begin{equation}
\label{eqI1}
V(r) = \left\{ \begin{array}{ll} 
-V_0 & , \; 0 \le r < a \\
0 & , \; r \ge a
\end{array} \right. \; ,
\end{equation}
where $V_0 > 0$ and $a > 0$. In order to find real-energy solutions, one normally matches
the wave function inside the well to a superposition of incoming and outgoing plane waves
outside the well (or to an exponentially damped function in the case of a bound state). Following 
Siegert \cite{Sieg39}, however, we require that the solution for $r\ge a$ is proportional to 
$\exp{(\mathrm{i} k r)}$, i.e., we apply the Siegert boundary condition
\begin{equation}
\label{eq4}
\left.\frac{\mathrm{d}}{\mathrm{d}r}\varphi(r)\right|_{r = a} = \mathrm{i} k \varphi(a) \; .
\end{equation}
If, in addition, we demand vanishing of the wave function at the origin, the wave number $k$ must
satisfy the following relation:
\begin{equation}
\label{eqI2}
\mathrm{i} k = \sqrt{k^2 + 2 V_0} \cot{(\sqrt{k^2 + 2 V_0}a)} \; .
\end{equation}

\begin{figure}
\includegraphics[width=9.5cm,origin=c,angle=0]{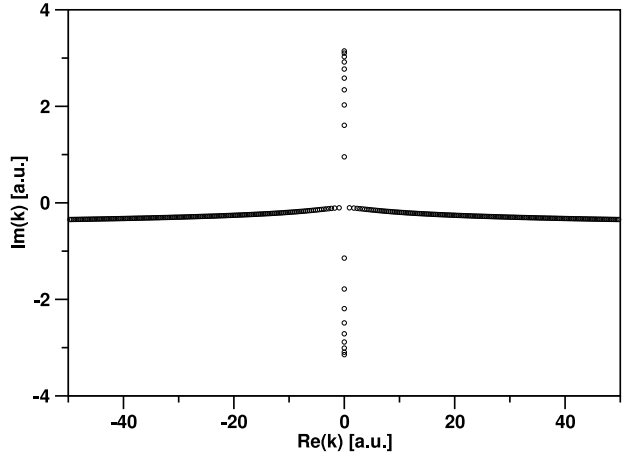}
\caption[]{The complex $k$ spectrum for a particle in the step potential of Eq. (\ref{eqI1}) with $V_0 = 5$ and $a=10$.
The spectrum is entirely discrete. Ten bound states
[$\mathrm{Re}(k) = 0$, $\mathrm{Im}(k) > 0$] are present, as well as nine antibound
[$\mathrm{Re}(k) = 0$, $\mathrm{Im}(k) < 0$]. The positive $\mathrm{Re}(k)$
branch of the spectrum shows the complex wave numbers of the outgoing Siegert pseudostates, the negative
$\mathrm{Re}(k)$ branch the incoming.}
\label{fig1}
\end{figure}

This transcendental equation can be solved for only a discrete set of $k$ values. As was done, for instance,
in Ref. \cite{KoMo84}, we solve Eq. (\ref{eqI2}) numerically, after setting $V_0=5$ and $a=10$. The resulting
discrete $k$ spectrum is presented in Fig. \ref{fig1}. There are $10$ bound states, which appear on the positive
imaginary axis. The nine antibound states lie on the negative imaginary axis. The solutions with $\mathrm{Re}(k)>0$
are associated with outgoing Siegert pseudostates. Similarly, states with $\mathrm{Re}(k)<0$ refer to incoming
Siegert pseudostates. 

Traditionally, Siegert-state theory has focused on scattering resonances and decaying states. See 
Refs. \cite{Newt02,MoGe73,ToOs97,ToOs98} for an overview of the literature. Several methods exist that allow
one to directly calculate the complex energy of a decaying state: complex scaling 
\cite{Rein82,Mois98,ReMc78}, complex absorbing potentials \cite{SaCe02,RiMe93}, and the direct solution of the 
Schr\"odinger equation subject to the Siegert boundary condition, Eq. (\ref{eq4}). This third approach used to be
numerically inefficient, because a nonlinear eigenvalue problem had to be solved in an iterative fashion
\cite{BaJu72,McRe79,Schn81}. 

A major step forward was made by Tolstikhin, Ostrovsky, and Nakamura \cite{ToOs97,ToOs98}. Their method provides, 
after solving a single generalized eigenvalue problem, access not only to the bound, antibound, and resonance states,
but also to the discretized pseudocontinuum. In addition, it becomes possible to derive fundamental properties of
Siegert pseudostates using simple and elegant mathematical techniques. An application of the method of 
Refs. \cite{ToOs97,ToOs98} to a molecular fragmentation problem is the subject of Ref. \cite{HaGr02}. In that work,
only bound states and outgoing Siegert pseudostates were utilized as fragmentation-channel basis functions.

A point of central concern in this paper is the time evolution of a wave packet expanded in terms of Siegert 
pseudostates. This aspect was first addressed by Yoshida {\em et al.} \cite{YoWa99}. They successfully 
introduced Siegert pseudostates as a basis capable of eliminating artificial boundary reflections.
In a subsequent paper \cite{TaWa01}, it is argued that the time evolution for $t > 0$ is given by 
\begin{equation}
\label{eq80}
\psi(r;t) = \sum_{n} (\varphi_n|\psi) \mathrm{e}^{-\mathrm{i} E_n t} \varphi_n(r) \; ,
\end{equation}
where the imaginary part of the complex energy $E_n$ is negative. $\varphi_n(r)$ is the spatial representation
of the $n$th Siegert pseudostate, and 
\begin{equation}
\label{eq49a}
(\varphi_n|\psi) = \int_0^a \varphi_n(r) \psi(r;t=0) \mathrm{d}r \; .
\end{equation}

In Sec. \ref{sec2}, we review some ideas of the formalism of Tolstikhin, Ostrovsky, and Nakamura. We extend their
work by providing arguments why certain subsets of Siegert pseudostates may be employed as bases. Moreover, we 
investigate time evolution in the context of Siegert pseudostates. We find that rigorous Siegert-pseudostate theory, 
in general, necessitates {\em nonexponential} time evolution, in contradiction to Eq. (\ref{eq80}). Numerical evidence 
is presented in Sec. \ref{sec3}. The calculations demonstrate that for a fast-moving wave packet Eq. (\ref{eq80}) 
is accurate, while its performance deteriorates as the average energy of the wave packet is decreased. 
Section \ref{sec4} concludes. Atomic units are used throughout.

\section{Mathematical considerations}
\label{sec2}

Consider the radial Schr\"odinger equation 
\begin{equation}
\label{eq1}
\hat{H}\varphi(r) = E \varphi(r) \; ,
\end{equation}
where 
\begin{equation}
\label{eq2}
\hat{H} = -\frac{1}{2} \frac{\mathrm{d}^2}{\mathrm{d}r^2} + V(r) \; .
\end{equation}
$V(r)$ is an arbitrary (short-range) potential. 
We seek solutions $\varphi(r)$ for $r \in [0,a]$ ($a>0$), such that at $r=0$ the 
usual boundary condition for the radial eigenvalue problem 
\begin{equation}
\label{eq3}
\varphi(0) = 0
\end{equation}
and at $r=a$ the {\em Siegert} boundary condition, Eq. (\ref{eq4}), 
are satisfied. The solutions $\varphi(r)$ are referred to as {\em Siegert pseudostates} \cite{ToOs97,ToOs98}.
Equation (\ref{eq4}) allows us to separate solutions to the Schr\"odinger equation 
into purely outgoing, purely incoming, bound, and antibound states. 

The wave number $k$ in Eq. (\ref{eq4}) is in general complex. Its relation to the eigenenergy $E$
is given by
\begin{equation}
\label{eq5}
E = \frac{k^2}{2} + V(a) \; ,
\end{equation}
which implies that $a$ should be chosen sufficiently large, if possible, such that $V(r) = \mathrm{const.}$ 
for $r \ge a$. Applying Eq. (\ref{eq5}) in the case of a long-range potential introduces the approximation
of forcing the potential to be constant beyond $r=a$, while preserving continuity of the potential at that 
point---in contrast to the approach taken in Ref. \cite{ToOs97,ToOs98}, which enforces a discontinuity by 
setting $V(a) = 0$ in Eq. (\ref{eq5}).

Utilizing a set of $N$ linearly independent, not necessarily orthogonal basis functions
$\{y_j(r): j=1,\ldots,N\}$ on the interval $[0,a]$ and assuming completeness in the limit 
$N \rightarrow \infty$, we can make the ansatz
\begin{equation}
\label{eq6}
\varphi(r) = \sum_{j=1}^N c_j y_j(r) \; .
\end{equation}
If we insert this into Eq. (\ref{eq1}), multiply from the left by $y_i(r)$, and integrate over $r$ from
$0$ to $a$, we find
\begin{widetext}
\begin{equation}
\label{eq7}
-\frac{1}{2}\int_0^a y_i(r) \frac{\mathrm{d}^2}{\mathrm{d}r^2} \sum_{j=1}^N c_j y_j(r) \mathrm{d}r 
 +\int_0^a y_i(r) V(r) \sum_{j=1}^N c_j y_j(r) \mathrm{d}r
= E \int_0^a y_i(r) \sum_{j=1}^N c_j y_j(r) \mathrm{d}r
\end{equation}
\end{widetext}
The term containing the second derivative with respect to $r$ can be integrated by parts. 
Thus, after applying the boundary conditions in Eqs. (\ref{eq3}) and (\ref{eq4}), we define
\begin{eqnarray}
\label{eq8}
\tilde{H}_{ij} & = & \int_0^a \frac{\mathrm{d}}{\mathrm{d}r} y_i(r) \frac{\mathrm{d}}{\mathrm{d}r}
y_j(r) \mathrm{d}r \\
& & + 2 \int_0^a y_i(r) [V(r)-V(a)] y_j(r) \mathrm{d}r \; , \nonumber \\
\label{eq9}
S_{ij} & = & \int_0^a y_i(r) y_j(r) \mathrm{d}r \; , \\
\label{eq10}
L_{ij} & = & y_i(a) y_j(a) \; , \\
\label{eq11}
\varkappa & = & \mathrm{i} k \; , 
\end{eqnarray}
and arrive at the nonlinear eigenvalue problem
\begin{equation}
\label{eq12}
\left(\tilde{\bm H} + \varkappa^2 {\bm S} - \varkappa {\bm L}\right){\bm c} = {\bm 0} \; .
\end{equation}
We refer to the eigenvector ${\bm c}$ in this equation as {\em Siegert pseudovector}.
In practical calculations, the basis functions $y_j(r)$ may be chosen to be real. Thus, the matrices
$\tilde{\bm H}$, ${\bm S}$, and ${\bm L}$ may be assumed to be symmetric and elements of 
$\mathbb{R}^{N \times N}$. Note, however, that the vector ${\bm c}$, consisting of the expansion
coefficients $c_j$, $j=1,\ldots,N$ [see Eq. (\ref{eq6})], is complex in general: 
${\bm c} \in \mathbb{C}^N$. If Eq. (\ref{eq12}) is satisfied by an eigenpair $(\varkappa,{\bm c})$, 
then---in view of the realness of $\tilde{\bm H}$, ${\bm S}$, and ${\bm L}$---the pair
$(\varkappa^{\ast},{\bm c}^{\ast})$ also forms a solution.

Tolstikhin, Ostrovsky, and Nakamura \cite{ToOs97,ToOs98} observed that the nonlinear eigenvalue problem 
of Eq. (\ref{eq12}) can be recast in terms of a linear one by defining
\begin{equation}
\label{eq13}
{\bm d} = \varkappa {\bm c} \; . 
\end{equation}
Then, Eq. (\ref{eq12}) goes over into
\begin{equation}
\label{eq14}
-\tilde{\bm H} {\bm c} = \varkappa \left({\bm S} {\bm d} - {\bm L} {\bm c}\right) \; ,
\end{equation}
and, trivially, 
\begin{equation}
\label{eq15}
{\bm S}{\bm d} = \varkappa {\bm S}{\bm c} \; .
\end{equation}
These two matrix equations are equivalent to the generalized eigenvalue problem
\begin{equation}
\label{eq16}
{\bm A}{\bm x} = \varkappa {\bm B}{\bm x} \; .
\end{equation}
The symmetric matrices ${\bm A}$ and ${\bm B}$ $\in \mathbb{R}^{2N \times 2N}$ are
given by
\begin{equation}
\label{eq17}
{\bm A} = \left[\begin{array}{cc}
-\tilde{\bm H} & {\bm 0} \\
{\bm 0}   & {\bm S}
\end{array}\right] \; ,
\end{equation}
\begin{equation}
\label{eq18}
{\bm B} = \left[\begin{array}{cc}
-{\bm L} & {\bm S} \\
{\bm S}   & {\bm 0}
\end{array}\right] \; ,
\end{equation}
and the eigenvector ${\bm x} \in \mathbb{C}^{2N}$ in Eq. (\ref{eq16}) is composed of the
two vectors ${\bm c}$ and ${\bm d}$:
\begin{equation}
\label{eq19}
{\bm x} = \left[\begin{array}{c}
{\bm c} \\
{\bm d} 
\end{array}\right] \; .
\end{equation}

Let us denote by $\tilde{N}$ the number of linearly independent solutions ${\bm x}_n$ 
(eigenvalue $\varkappa_n$) to the generalized eigenvalue problem in Eq. (\ref{eq16}).
(It is by no means clear that $\tilde{N} = 2N$.) We then define the matrices
\begin{equation}
\label{eq20}
{\bm X} = [{\bm x}_1,\ldots,{\bm x}_{\tilde{N}}] \in \mathbb{C}^{2N \times \tilde{N}}
\end{equation}
and 
\begin{equation}
\label{eq21}
{\bm K} = \mathrm{diag}(\varkappa_1,\ldots,\varkappa_{\tilde{N}}) \in \mathbb{C}^{\tilde{N} \times \tilde{N}} \; ,
\end{equation}
such that 
\begin{equation}
\label{eq22}
{\bm A}{\bm X} = {\bm B}{\bm X}{\bm K} \; .
\end{equation}

The overlap matrix ${\bm S}$ [Eq. (\ref{eq9})] is positive-definite and therefore invertible.
Its inverse, ${\bm S}^{-1}$, is also symmetric. It can be concluded that the inverse of ${\bm B}$
exists, since it is easily seen by direct construction that 
\begin{equation}
\label{eq23}
{\bm B}^{-1} = \left[\begin{array}{cc}
{\bm 0}      & {\bm S}^{-1} \\
{\bm S}^{-1} & {\bm S}^{-1}{\bm L}{\bm S}^{-1}
\end{array}\right] \; .
\end{equation}
The real symmetric matrix ${\bm B}$ can be diagonalized via an orthogonal similarity transformation,
\begin{equation}
\label{eq24}
{\bm U}^{\mathrm{T}}{\bm B}{\bm U} = {\bm D} \; ,
\end{equation}
where ${\bm U} \in \mathbb{R}^{2N \times 2N}$ is orthogonal, ${\bm U}^{\mathrm{T}}{\bm U} = {\bm \openone}$,
and ${\bm D} \in \mathbb{R}^{2N \times 2N}$ is diagonal. All diagonal elements of ${\bm D}$ differ from 
zero, for ${\bm D}$ is similar to ${\bm B}$ and thus also invertible. With this in mind, the matrix
\begin{equation}
\label{eq25}
\tilde{\bm U} = {\bm U}{\bm D}^{-1/2} \in \mathbb{C}^{2N \times 2N}
\end{equation}
can be introduced in a meaningful way. While ${\bm B}$ is invertible [Eq. (\ref{eq23})], it is in 
general neither positive- nor negative-definite. Hence, some of the columns of $\tilde{\bm U}$ 
are real, but the others are purely imaginary. The matrix $\tilde{\bm U}$ can be utilized to convert the 
real symmetric, indefinite generalized eigenvalue problem, Eq. (\ref{eq22}), to a standard one:
\begin{equation}
\label{eq26}
\tilde{\bm A}\tilde{\bm X} = \tilde{\bm X}{\bm K} \; .
\end{equation}
Here,
\begin{equation}
\label{eq27}
\tilde{\bm A} = \tilde{\bm U}^{\mathrm{T}}{\bm A}\tilde{\bm U}
\end{equation}
and 
\begin{equation}
\label{eq28}
\tilde{\bm X} = {\bm D}\tilde{\bm U}^{\mathrm{T}}{\bm X} \; .
\end{equation}

The matrix $\tilde{\bm A}$ is complex symmetric. Unfortunately, complex symmetry is not a very useful
property, as it is known that {\em any} complex matrix of square format is similar to a complex symmetric 
matrix (see, for instance, Ref. \cite{SaCe02}). Generally, it is not possible to guarantee diagonalizability, 
i.e. the existence of a basis of eigenvectors, of a complex symmetric matrix. Note that $\tilde{\bm A}$ 
consists of purely real and purely imaginary matrix blocks, but whether this helps to prove its 
diagonalizability is currently unclear. A sufficient condition for diagonalizability is that all $2N$ 
eigenvalues $\varkappa_n$ are distinct. Degeneracies could cause difficulties, but in numerical calculations 
true (accidental) degeneracies are practically never encountered. Under the assumption that 
$\tilde{\bm A}$ {\em is} diagonalizable, $\tilde{N} = 2N$ and the matrix  
\begin{equation}
\label{eq29}
\tilde{\bm X} = [\tilde{\bm x}_1,\ldots,\tilde{\bm x}_{2N}] \in \mathbb{C}^{2N \times 2N}
\end{equation}
may be chosen to be complex orthogonal, 
\begin{equation}
\label{eq30}
\tilde{\bm X}^{\mathrm{T}}\tilde{\bm X} = {\bm \openone} \; , 
\end{equation}
as shown, e.g., in Ref. \cite{SaCe02}. The $2N$ vectors $\tilde{\bm x}_n$ are linearly independent and, 
consequently, form a basis of $\mathbb{C}^{2N}$. This fact can be expressed in compact form in terms of 
the completeness relation 
\begin{equation}
\label{eq31}
\sum_{n=1}^{2N} \tilde{\bm x}_n \tilde{\bm x}_n^{\mathrm{T}} = {\bm \openone} \; .
\end{equation}

From Eqs. (\ref{eq30}) and (\ref{eq31}) it follows for the solutions ${\bm x}_n$ of the original generalized
eigenvalue problem in Eq. (\ref{eq16}) that
\begin{equation}
\label{eq32}
{\bm x}_m^{\mathrm{T}}{\bm B}{\bm x}_n = \delta_{mn} \; , \; m,n=1,\ldots,2N \; ,
\end{equation}
and
\begin{equation}
\label{eq33}
\sum_{n=1}^{2N} {\bm x}_n {\bm x}_n^{\mathrm{T}} = {\bm B}^{-1} \; .
\end{equation}
Combining Eq. (\ref{eq32}) with Eqs. (\ref{eq13}), (\ref{eq18}), and (\ref{eq19}), and employing the
normalization convention of Refs. \cite{ToOs97,ToOs98}, the orthonormality relation satisfied by the 
eigenvectors ${\bm c}_n$ of the nonlinear eigenvalue problem in Eq. (\ref{eq12}) reads
\begin{equation}
\label{eq34}
{\bm c}_m^{\mathrm{T}}{\bm S}{\bm c}_n - \frac{{\bm c}_m^{\mathrm{T}}{\bm L}{\bm c}_n}{\varkappa_m + \varkappa_n}
= \delta_{mn} \; , \; m,n=1,\ldots,2N \; .
\end{equation}
Take notice of the occurrence of a singularity in this expression if $\varkappa_m + \varkappa_n$
approaches $0$ for some pair of indices $m,n$. This happens whenever there is a deeply bound state 
(eigenvalue $\varkappa_m$), since there exists a corresponding antibound state whose eigenvalue
$\varkappa_n$ equals $-\varkappa_m$ to machine precision.

Even in the absence of the term involving the surface matrix ${\bm L}$, Eq. (\ref{eq34}) could not be used
to establish a proper inner product on $\mathbb{C}^N$, because, for a given vector 
${\bm v}\in\mathbb{C}^N\setminus\{{\bm 0}\}$, ${\bm v}^{\mathrm{T}}{\bm S}{\bm v}$ is not necessarily different
from zero. The occurrence of indefinite inner products is also well known in the context of complex scaling
\cite{Rein82,Mois98} and in the method of complex absorbing potentials \cite{SaCe02,RiMe93}. The following 
three relations are consequences of Eqs. (\ref{eq13}), (\ref{eq19}), (\ref{eq23}), and (\ref{eq33}):
\begin{eqnarray}
\label{eq35}
\sum_{n=1}^{2N} \frac{1}{\varkappa_n}{\bm c}_n {\bm c}_n^{\mathrm{T}} & = & {\bm 0} \; , \\
\label{eq36}
\sum_{n=1}^{2N} {\bm c}_n {\bm c}_n^{\mathrm{T}} & = & 2 {\bm S}^{-1} \; , \\
\label{eq37}
\sum_{n=1}^{2N} \varkappa_n {\bm c}_n {\bm c}_n^{\mathrm{T}} & = & 2 {\bm S}^{-1} {\bm L} {\bm S}^{-1} \; . 
\end{eqnarray}
The normalization suggested by Eq. (\ref{eq34}) has been applied.

Equation (\ref{eq36}) demonstrates that any vector ${\bm v}\in\mathbb{C}^N$ can be represented as a 
superposition of the Siegert pseudovectors ${\bm c}_n$:
\begin{equation}
\label{eq38}
{\bm v} = \frac{1}{2}\sum_{n=1}^{2N} \left({\bm c}_n^{\mathrm{T}}{\bm S}{\bm v}\right){\bm c}_n \; .
\end{equation}
This representation, however, is not unique, for the vectors ${\bm c}_n$, $n=1,\ldots,2N$, form an overcomplete
subset of $\mathbb{C}^N$. The rank of the matrix 
${\bm C} = [{\bm c}_1,\ldots,{\bm c}_{2N}] \in \mathbb{C}^{N \times 2N}$ cannot be greater than $N$ [in fact,
because of Eq. (\ref{eq36}), $\mathrm{rank}({\bm C}) = N$]. Thus, the vectors ${\bm c}_1,\ldots,{\bm c}_{2N}$
are linearly dependent, i.e., the equation
\begin{equation}
\label{eq39}
\sum_{n=1}^{2N} \alpha_n {\bm c}_n = {\bm 0}
\end{equation}
can be satisfied by some $\alpha_n \ne 0$. If we define 
\begin{equation}
\label{eq40}
M_{mn} = {\bm c}_m^{\mathrm{T}}{\bm S}{\bm c}_n \; ,
\end{equation}
then 
\begin{equation}
\label{eq41}
\sum_{n=1}^{2N} M_{mn} \alpha_n = 0 \; , m=1,\ldots,2N \; ,
\end{equation}
from which we may conclude that the matrix ${\bm M} \in \mathbb{C}^{2N \times 2N}$ cannot be invertible 
(since not all $\alpha_n$ have to be equal to $0$). The nonuniqueness of the representation in Eq. (\ref{eq38})
is directly linked to the noninvertibility of ${\bm M}$.

Another instructive way of looking at this is to make use of Eq. (\ref{eq36}) to derive the following relation:
\begin{equation}
\label{eq42}
{\bm M} {\bm M} = 2 {\bm M} \; .
\end{equation}
If ${\bm M}$ were invertible, then we would have ${\bm M} = 2 {\bm \openone}$. This contradicts the 
orthonormality relation, Eq. (\ref{eq34}):
\begin{equation}
\label{eq43}
M_{mn} = \delta_{mn} + \frac{{\bm c}_m^{\mathrm{T}}{\bm L}{\bm c}_n}{\varkappa_m + \varkappa_n} \; .
\end{equation}
For example, for $m=n$ the surface term can be written as $\varphi_m(a)^2/2\varkappa_m$ [Eqs. (\ref{eq6})
and (\ref{eq10})], which differs from unity in general.

In Appendix \ref{appA}, we demonstrate that a subset of $\{{\bm c}_1,\ldots,{\bm c}_{2N}\}$ can be found,
comprising exactly $N$ vectors and forming a basis of $\mathbb{C}^N$. Without loss of generality, the elements
of this subset are taken to be the vectors ${\bm c}_1,\ldots,{\bm c}_{N}$. The proper completeness relation 
allows one to represent an arbitrary vector ${\bm v} \in \mathbb{C}^N$ in a unique way:
\begin{equation}
\label{eq44}
{\bm v} = \sum_{n=1}^N \alpha_n {\bm c}_n \; .
\end{equation}
Necessarily, 
\begin{equation}
\label{eq45}
\sum_{n=1}^{N} M_{mn} \alpha_n = {\bm c}_m^{\mathrm{T}}{\bm S}{\bm v} \; , m=1,\ldots,N \; .
\end{equation}
Note that the matrix elements $M_{mn}$ in this equation refer only to the selected subset.
The linear independence of the Siegert pseudovectors ${\bm c}_1,\ldots,{\bm c}_N$ ensures the 
invertibility of ${\bm M} \in \mathbb{C}^{N \times N}$ [cf. Eqs. (\ref{eq39}), (\ref{eq40}), and (\ref{eq41})].
The expansion coefficients $\alpha_m$ are unique and are given by
\begin{equation}
\label{eq46}
\alpha_m = \sum_{n=1}^{N} \left({\bm M}^{-1}\right)_{mn} {\bm c}_n^{\mathrm{T}}{\bm S}{\bm v} \; .
\end{equation}
The associated completeness relation reads
\begin{equation}
\label{eq47}
\sum_{m=1}^{N} \sum_{n=1}^{N} {\bm c}_m \left({\bm M}^{-1}\right)_{mn} {\bm c}_n^{\mathrm{T}} = {\bm S}^{-1} \; .
\end{equation}

Let us now turn our attention to the problem of describing the time evolution of an outgoing wave packet utilizing
Siegert pseudostates. For that purpose, it is natural to choose for the basis $\{{\bm c}_1,\ldots,{\bm c}_N\}$
all outgoing Siegert pseudovectors [$\mathrm{Re}(k_n) > 0$] plus a complementary number of bound eigenvectors. 
(We provide in Sec. \ref{sec3} numerical evidence that these form indeed a basis of $\mathbb{C}^N$.)
Consider an initial wave packet $\psi(r;t=0)$ that is entirely localized within the interval $[0,a]$. Using
Eqs. (\ref{eq6}) and (\ref{eq47}), and assuming pure exponential time evolution for each Siegert pseudostate, 
the time evolution of the wave packet for $t>0$ is obtained from 
\begin{equation}
\label{eq48a}
\psi(r;t) = \sum_{m=1}^{N} \sum_{n=1}^{N} \left({\bm M}^{-1}\right)_{mn} (\varphi_n|\psi)
\mathrm{e}^{-\mathrm{i} E_m t} \varphi_m(r) \; ,
\end{equation}
where
\begin{equation}
\label{eq50a}
E_m = \frac{k_m^2}{2} + V(a) \; .
\end{equation}
One disadvantage of Eq. (\ref{eq48a}) is the need to invert a matrix. In addition, as we will see below,
the assumption of exponential time evolution is wrong in general for Siegert pseudostates.

A practically and formally more acceptable expression for the time evolution of the wave packet can 
be derived from the Mittag-Leffler partial fraction decomposition \cite{NaNi01} of the outgoing-wave Green's 
function represented with respect to Siegert pseudostates. This representation has been given in 
Refs. \cite{MoGe73,ToOs97,ToOs98}. The outgoing-wave Green's function satisfies the equations
\begin{eqnarray}
\label{eq51a}
(E-\hat{H})G(r,r';k) & = & \delta(r-r') \; , \\
\label{eq52a}
G(0,r';k) & = & 0 \; , \\
\label{eq53a}
\left.\frac{\mathrm{d}}{\mathrm{d}r}G(r,r';k)\right|_{r = a} & = & \mathrm{i} k G(a,r';k) \; ,
\end{eqnarray}
for $r,r' \in [0,a]$. These conditions can be translated to
\begin{equation}
\label{eq54a}
\left(\tilde{\bm H} + \varkappa^2 {\bm S} - \varkappa {\bm L}\right){\bm G}(k) = -2{\bm \openone} \; ,
\end{equation}
where the matrix representation of the outgoing-wave Green's function in the basis of the functions 
$y_j(r)$ is defined via the relation 
\begin{equation}
\label{eq55a}
G(r,r';k) = \sum_{i=1}^N \sum_{j=1}^N G_{ij}(k) y_i(r) y_j(r') \; .
\end{equation}
Making the ansatz
\begin{equation}
\label{eq56a}
{\bm G}(k) = \sum_{n=1}^{2N} \alpha_n {\bm c}_n {\bm c}_n^{\mathrm{T}} 
\end{equation}
and putting Eqs. (\ref{eq35}), (\ref{eq36}) to use, it follows from Eq. (\ref{eq54a}) that
\begin{equation}
\label{eq57a}
\alpha_n = \frac{1}{k_n(k-k_n)} \; .
\end{equation}
Hence,
\begin{equation}
\label{eq58a}
G(r,r';k) = \sum_{n=1}^{2N} \frac{\varphi_n(r) \varphi_n(r')}{k_n(k-k_n)} \; , \; r,r' \in [0,a] \; ,
\end{equation}
serves as an approximation to the outgoing-wave Green's function, within the framework of
the underlying finite basis set $\{y_j(r): j=1,\ldots,N\}$.

\begin{figure}
\includegraphics[width=4.5cm,origin=c,angle=0]{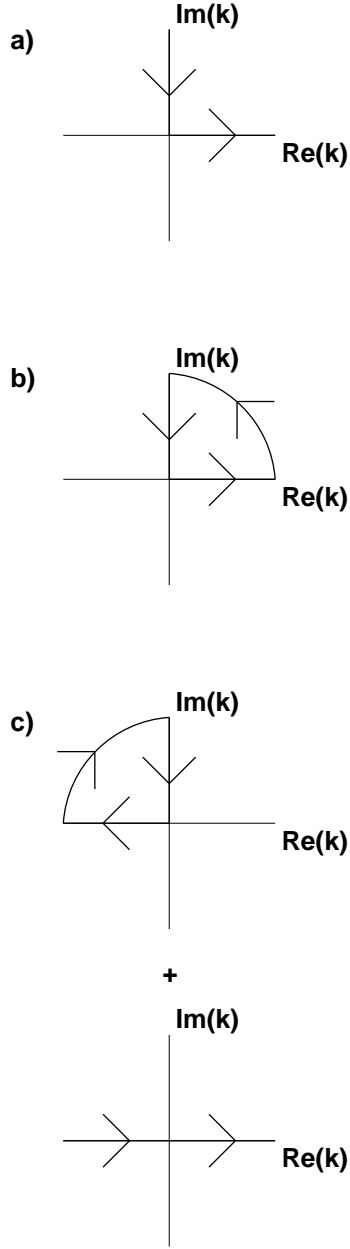}
\caption[]{a) The integration contour in the complex $k$ plane employed to evaluate Eq. (\ref{eq60}).
The integration proceeds along the imaginary axis from $\mathrm{i}\infty$ to the origin, and from there along the
real axis to $+\infty$. This contour corresponds to integration from $-\infty$ to $+\infty$ on the physical sheet
of the energy Riemann surface. b) If $t<0$, then the contour may be closed in the first quadrant of the $k$ plane.
c) The integration path
in a) may be evaluated for $t>0$ using one closed contour and one that runs from $-\infty$ to $+\infty$.}
\label{fig2}
\end{figure}

The Green's function in Eq. (\ref{eq58a}) allows us to determine the time evolution of the wave packet for $t>0$
\cite{GoWa64}:
\begin{widetext}
\begin{equation}
\label{eq59}
\psi(r;t) = \frac{\mathrm{i}}{2\pi} \int_{-\infty}^{\infty} \mathrm{d}E \, \mathrm{e}^{-\mathrm{i}Et}
\int_0^a \mathrm{d}r' G(r,r';k) \psi(r';t=0) 
= \sum_{n=1}^{2N} \beta_n(t) (\varphi_n|\psi) \varphi_n(r) \; .
\end{equation}
\end{widetext}
Here,
\begin{equation}
\label{eq60}
\beta_n(t) = \frac{\mathrm{i}}{2\pi} \int_{-\infty}^{\infty} \frac{\mathrm{e}^{-\mathrm{i}Et}}{k_n(k-k_n)}
\mathrm{d}E \; .
\end{equation}
The integration path lies on the physical sheet \cite{Newt02} of the energy Riemann surface, 
and runs from $-\infty$ to $+\infty$ infinitesimally above the real $E$ axis \cite{GoWa64}. In order to evaluate 
$\beta_n(t)$, we use Eq. (\ref{eq5}) and replace the integration path in the energy plane by the integration 
contour in the complex $k$ plane illustrated in Fig. \ref{fig2}a. For $t<0$, this contour can be closed in the 
first quadrant of 
the $k$ plane without affecting $\beta_n(t)$. The resulting integration loop is indicated in Fig. \ref{fig2}b. 
The integration path along the positive imaginary axis is infinitesimally displaced toward positive real $k$ values,
so that the integrand in Eq. (\ref{eq60}) is analytic on and inside the loop. Hence, $\beta_n(t)=0$ for $t<0$, as 
expected for a description based on the outgoing-wave Green's function. 

If $t>0$, then $\beta_n(t)$ can be written as the sum of two separate contour integrals. The respective contours
are shown in Fig. \ref{fig2}c. Thus we have 
\begin{equation}
\label{eq61}
\beta_n(t) = \beta_n^{(I)}(t) + \beta_n^{(II)}(t) \; ,
\end{equation}
where
\begin{equation}
\label{eq62}
\beta_n^{(I)}(t) = \left\{ \begin{array}{ll} 
\mathrm{e}^{-\mathrm{i} E_n t} & \textrm{for bound states}\\
0 & \textrm{otherwise}
\end{array} \right.
\end{equation}
and 
\begin{eqnarray}
\beta_n^{(II)}(t) & = & \frac{\mathrm{i}}{2\pi} 
\frac{\exp{\left\{-\mathrm{i}V(a)t\right\}}}{k_n}
\int_{-\infty}^{\infty} \frac{k}{k-k_n} \mathrm{e}^{-\mathrm{i} k^2 t/2} \mathrm{d}k \nonumber \\
& = & 
\label{eq64}
\frac{\mathrm{i}}{2\pi} \exp{\left\{-\mathrm{i}V(a)t\right\}} \\
& & \times \left\{ \frac{\mathrm{e}^{-\mathrm{i} \pi / 4}}{k_n}\sqrt{\frac{2\pi}{t}}
+ \int_{-\infty}^{\infty} \frac{\mathrm{e}^{-\mathrm{i} k^2 t/2}}{k-k_n} \mathrm{d}k \right\} \; . \nonumber 
\end{eqnarray}
From Eq. (\ref{eq35}) it follows that 
\begin{equation}
\label{eq65}
\sum_{n=1}^{2N} \frac{1}{k_n} (\varphi_n|\psi) \varphi_n(r) = 0 \; .
\end{equation}
Therefore, the term proportional to $t^{-1/2}$ in $\beta_n^{(II)}(t)$ does not contribute to 
$\psi(r;t)$, Eq. (\ref{eq59}), and we will not consider it further.
As demonstrated in Appendix \ref{appB}, the integral in Eq. (\ref{eq64}) can be expressed utilizing
the {\em Faddeeva} function \cite{FaTe61,FrCo61,Huml79,NeAg81,PoWi90,SuTh91,Well99}
\begin{equation}
\label{eq66}
w(z) = \frac{\mathrm{i}}{\pi} \int_{-\infty}^{\infty} \frac{\mathrm{e}^{-s^2}}{z-s} \mathrm{d}s \; , \;
\mathrm{Im}(z) > 0 \; .
\end{equation}
Hence, on combining Eqs. (\ref{eq61}), (\ref{eq62}), (\ref{eq64}), and the results from Appendix \ref{appB},
the time-evolution coefficients $\beta_n(t)$, $t>0$, read
\begin{widetext}
\begin{equation}
\label{eq67}
\beta_n(t) = \exp{\left\{-\mathrm{i}V(a)t\right\}} 
 \times \left\{ \begin{array}{ll}
\mathrm{e}^{-\mathrm{i} k_n^2 t/2} - \frac{1}{2} w\left(\mathrm{e}^{\mathrm{i} \pi / 4}\sqrt{\frac{t}{2}}k_n\right)
& \; \textrm{for bound or outgoing states}\\
\\
\frac{1}{2} w\left(-\mathrm{e}^{\mathrm{i} \pi / 4}\sqrt{\frac{t}{2}}k_n\right) 
& \; \textrm{for antibound or incoming states}
\end{array} \right.
\; . 
\end{equation}
\end{widetext}
(The square root in the argument of the Faddeeva function must be interpreted as a positive number.)
The result in Eq. (\ref{eq67}), together with Eq. (\ref{eq59}), represents---within the framework of 
Siegert pseudostates---the most consistent approach to describing the time evolution of a wave packet. 
It is interesting to observe that in the limit $t\rightarrow 0^+$, all $\beta_n(t)$ tend to $1/2$, for
the Faddeeva function goes to unity in this limit. This ensures, in view of Eq. (\ref{eq36}), that
$\lim_{t\rightarrow 0^+} \psi(r;t) = \psi(r;t=0)$. Equations (\ref{eq59}) and (\ref{eq67}) 
also allow us to determine the long-term behavior of the wave packet in the interval $[0,a]$. Using
the asymptotic expansion of the Faddeeva function \cite{AbSt70} and Eq. (\ref{eq65}), the wave packet
for large $t$ is seen to consist of bound-state components evolving according to the 
$\exp{\left\{-\mathrm{i} E_n t\right\}}$ factor in Eq. (\ref{eq67}) as well as a decaying component 
that evolves in time as $t^{-3/2}$. This specific power-law behavior of a decaying state---in the long-time 
limit---is in fact well known \cite{MoGe73,GoWa64,JaSa61,Wint61,DiRe02} (see Refs. \cite{NiBe77,Nico02}
for an alternative view).

\section{Numerical Studies}
\label{sec3}

In this section, we apply the techniques developed so far to a wave packet in the model potential of Eq. (\ref{eqI1})
[$V_0=5$ and $a=10$]. For the basis $\{y_j(r): j=1,\ldots,N\}$, a finite-element basis set based on fifth-order 
Hermite interpolating polynomials was used \cite{Bath76,BaWi76,BrSc93,AcSh96,ReBa97,MeGr97,SaCh04}. A mesh of 
evenly-spaced nodes was defined radially, with three finite elements centered at the $i$th node, $r_i$. All three 
vanish in $[0,r_{i-1}]$ and $[r_{i+1},a]$. Their behavior at $r_i$ separates them into three types. The zero-type 
polynomials have a finite function value at the $i$th node, 
but zero first and second derivatives. The one-type have a finite first derivative, but zero function value and 
zero second derivative. The two-type have a finite curvature, but vanishing zero and first derivatives.  In 
order to enforce vanishing at the origin, Eq. (\ref{eq3}), only the one- and two-types are considered at the 
first node (at $r=0$). Thus, the dimension of the finite-element basis set ($N$) is equal to one less than three 
times the number of nodes.

We numerically \cite{Lapa99} solve the generalized eigenvalue problem of Eq. (\ref{eq16}) to obtain $2N$ Siegert 
pseudostates with $2N$ distinct eigenvalues $\varkappa_n$.  Appendix \ref{appA} proves that certain subsets of $N$ 
Siegert pseudostates achieve completeness in $\mathbb{C}^N$.  However, which $N$ states must be selected is not clear.  
By looking at the eigenvalues of the matrix ${\bm M} \in \mathbb{C}^{N \times N}$ [Eq. (\ref{eq40})], one can determine 
whether or not a selected subset of $N$ vectors is complete. If the vectors form a basis, the ${\bm M}$ matrix will 
have $N$ (not necessarily distinct) eigenvalues which differ from zero. For the potential under consideration, the 
number of bound states equals the number of antibound states. (In the Introduction, we mentioned that there are
$10$ physical bound and nine physical antibound states. The $10$th antibound state found in the numerical calculation
cannot be converged and does not correspond to an eigenstate of the Hamiltonian.) All combinations of bound or 
antibound with incoming or outgoing Siegert pseudostates supply $N$ distinct vectors, giving four plausible choices for 
a complete, useful basis set. The eigenvalues of ${\bm M}$ were calculated for each case, and in so far as zero was 
never one of them, all 
combinations proved to span $\mathbb{C}^N$. More specifically, more than $95$ \%  of all eigenvalues of 
${\bm M}$ equal unity to machine precision. The rest have a magnitude ranging from $0.93$ to $230$. 

To assess the ability of the Siegert pseudostates to represent a wave packet at $t=0$, it is first necessary to 
investigate the limiting accuracy of the underlying finite-element basis set. An initial wave packet is given the 
form of a Gaussian multiplied by a plane wave:
\begin{equation}
\label{eq68}
\psi(r;t=0) = \exp{[-\frac{\left(r-r_0\right)^2}{2\xi^2} + \mathrm{i} k_0 \left(r-r_0\right)]} \; .
\end{equation}
The Gaussian is centered in the middle of the well at $r_0 = 5$, and $\xi=0.5$ is chosen to ensure that contribution 
outside the interval $[0,a]$ is negligible. 

We represent the wave packet as a superposition of the finite elements,
\begin{equation}
\label{eq69}
\psi(r;t=0) \approx \sum_{j=1}^{N} \alpha_j y_j(r) \; ,
\end{equation}
where the expansion coefficients are chosen to minimize 
\begin{equation}
\label{eq70}
\chi_1^{2} = \int_{0}^{a} \left|\psi(r;t=0)-\sum_{j=1}^{N}\alpha_j y_j(r)\right|^2 \mathrm{d}r \; ,
\end{equation}
i.e.,
\begin{equation}
\label{eq71}
\alpha_i = \sum_{j=1}^{N} \left({\bm S}^{-1}\right)_{ij} \int_{0}^{a} y_j(r) \psi(r;t=0) \mathrm{d}r \; .
\end{equation}
On the basis of $\chi_1^{2}$, we determine how accurately our finite set of piecewise-defined polynomials is able 
to approximate a specific element of the infinite-dimensional Hilbert space.

Let us next consider the completeness relation expressed by Eq. (\ref{eq47}). The set of $N$ selected Siegert 
pseudostates must be capable of representing the unique coefficient vector 
${\bm \alpha} \in \mathbb{C}^N$,
\begin{equation}
\label{eq72}
\tilde{\bm \alpha} = \sum_{n=1}^{N} \gamma_n {\bm c}_n \; ,
\end{equation}
where, ideally, $\tilde{\bm \alpha} = {\bm \alpha}$. According to Eq. (\ref{eq47}), 
\begin{equation}
\label{eq73}
\gamma_m = \sum_{n=1}^N \left({\bm M}^{-1}\right)_{mn} {\bm c}_n^{\mathrm{T}} {\bm S} {\bm \alpha} \; .
\end{equation}
The quality of the Siegert pseudovector expansion is measured by 
\begin{equation}
\label{eq74}
\chi_2^{2} = \sum_{j=1}^{N}|\alpha_j - \tilde{\alpha}_j|^2 \; .
\end{equation}
Now with the expansion coefficients themselves superpositions of the Siegert pseudostates, we calculate 
\begin{equation}
\label{eq75}
\chi_3^{2} = \int_{0}^{a} \left|\psi(r;t=0)-\sum_{n=1}^{N} \gamma_n \varphi_n(r)\right|^2 \mathrm{d}r 
\end{equation}
to test the ability of the chosen subset of $N$ Siegert pseudostates to accurately reproduce the wave packet at $t=0$.  

We now focus on Eq. (\ref{eq36}), which allows us to expand the wave packet in terms of all $2N$ Siegert pseudostates:
\begin{equation}
\label{eq76}
\chi_4^{2} = \sum_{j=1}^{N}\left|\alpha_j - \sum_{n=1}^{2N} \zeta_n c_{jn}\right|^2 \; ,
\end{equation}
\begin{equation}
\label{eq77}
\zeta_n = \frac{1}{2} {\bm c}_n^{\mathrm{T}} {\bm S} {\bm \alpha} \; ,
\end{equation}
\begin{equation}
\label{eq78}
\chi_5^{2} = \int_{0}^{a} \left|\psi(r;t=0)-\sum_{n=1}^{2N} \zeta_n \varphi_n(r)\right|^2 \mathrm{d}r \; .
\end{equation}

\begin{table}
\caption[]{The $\chi^2$'s described in Eqs. (\ref{eq70})-(\ref{eq78}), as a function of finite-element basis size, $N$,
calculated for the case where $k_0=15$ in Eq. (\ref{eq68}). $\chi_1^2$ demonstrates convergence to accurate wave-packet
reproduction as
the number of finite-element basis functions is increased. $\chi_2^2$ and $\chi_4^2$ show the ability of the
expansions of Eqs. (\ref{eq48a}) and (\ref{eq59}), respectively, to match the unique coefficients of Eq. (\ref{eq71}).
$\chi_3^2$ and $\chi_5^2$ confirm that these expansions are successful in reproducing the initial wave packet at the
level limited by the underlying finite elements. Notation $x[y]$ stands for $x \times 10^y$.}
\label{tab1}
\begin{ruledtabular}
\begin{tabular}{cccccc}
$N$  & $\chi_1^{2}$ & $\chi_2^{2}$ & $\chi_3^{2}$ & $\chi_4^{2}$ & $\chi_5^{2}$ \\
\hline
$20$  & $0.89$ & $2.7[-19]$ & $0.89$ & $9.6[-20]$ & $0.89$ \\
$80$  & $ 1.5[-3]$ & $1.4[-17]$ & $1.5[-3]$ & $2.2[-19]$ & $1.5[-3]$ \\
$200$  & $2.0[-7]$ & $4.5[-20]$ & $2.0[-7]$ & $5.3[-21]$ & $2.0[-7]$\\
$380$  & $3.0[-11]$ & $7.8[-20]$ & $3.0[-11]$ & $9.8[-21]$ & $3.0[-11]$\\
$620$  & $5.5[-14]$ & $4.5[-18]$ & $5.5[-14]$ & $9.9[-19]$ & $5.5[-14]$\\
\end{tabular}
\end{ruledtabular}
\end{table}

All five $\chi^{2}$'s were calculated as a function of $N$, the number of finite-element basis functions, with 
$k_0=15$ [see Eq. (\ref{eq68})]. The values are displayed in Table \ref{tab1}. Only $\chi_2^{2}$'s and $\chi_3^{2}$'s 
using bound and outgoing Siegert pseudostates are tabulated, since these states provided the qualitatively best  
results for time evolution based on Eq. (\ref{eq48a}). Other combinations of $N$ Siegert pseudostates will not be 
discussed further.

One observes in Table \ref{tab1} that as the number of basis functions being used is increased, $\chi_1^{2}$ 
approaches convergence. For all values of $N$, the high accuracy of $\chi_2^{2}$ confirms that the Siegert 
pseudovectors indeed form a basis for $\mathbb{C}^N$. Because of the excellent precision achieved at the vector 
level, we see that $\chi_3^{2}$, which measures the representability of the initial wave packet in terms of the 
spatial representation of the $N$ selected Siegert pseudostates, is only limited by the degree to which the finite 
elements are complete. The quantities $\chi_4^{2}$ and $\chi_5^{2}$ in Table \ref{tab1} demonstrate that similar
statements hold with regard to accuracy if one utilizes the overcomplete set of $2N$ Siegert pseudostates.

Now consideration is given to the propagation of a wave packet in time. To that end, we require a benchmark of 
comparison. Positive-energy eigensolutions in the interval $[0,a]$ were found for the potential of study, 
Eq. (\ref{eqI1}), by matching at the discontinuity to energy-normalized continuum solutions of the form 
$\sqrt{\frac{2}{\pi k}}\sin{(k r + \delta)}$. These were used, together with the bound-state solutions, to expand the 
Gaussian wave packet of Eq. (\ref{eq68}). We refer to this expansion as the ``analytical solution,'' even though
it should be pointed out that we carried out the integration over the energy continuum numerically. The validity of 
the analytical solution was established by comparison with the closed-form expression in Eq. (\ref{eq68}) at $t=0$. 
Numerical convergence of the energy integration was tested carefully and ensured.
Standard exponential time dependence was introduced for $t>0$. 

\begin{figure}
\includegraphics[width=9.5cm,origin=c,angle=0]{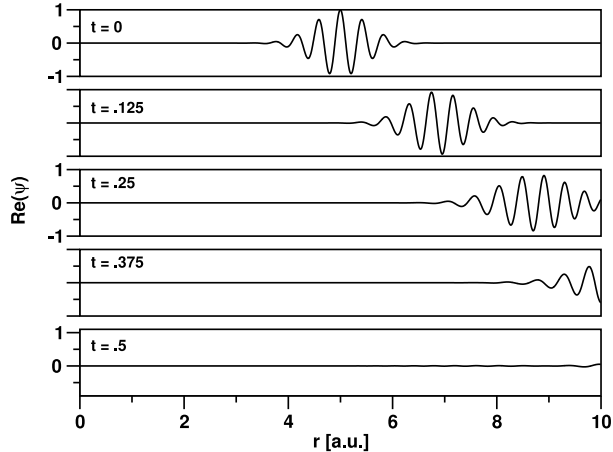}
\caption[]{Time evolution of the wave packet described at $t=0$ by Eq. (\ref{eq68}); $k_0=15$. In the $r$ interval
between $0$ and $10$, the potential is constant [$V_0=5$ and $a=10$ in Eq. (\ref{eqI1})].}
\label{fig3}
\end{figure}

\begin{figure}
\includegraphics[width=9.5cm,origin=c,angle=0]{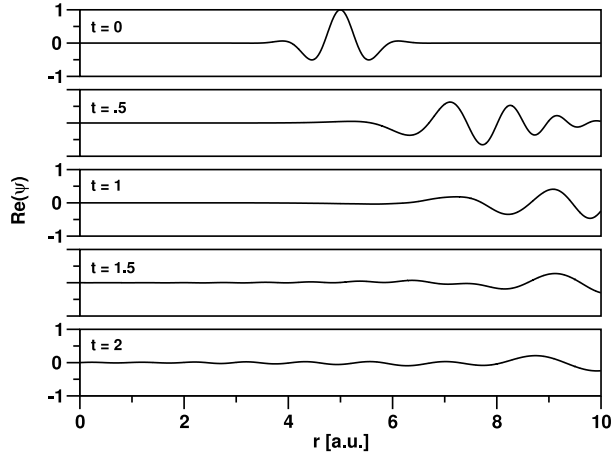}
\caption[]{The same parameters as in Fig. \ref{fig3}, but with $k_0=5$. Notice that there are significant
physical reflections as soon as the wave packet hits the discontinuity at $r=a$.}
\label{fig4}
\end{figure}

Once it is determined that the analytical solution gives a wave packet representation that is exact to machine precision
(at $t=0$), we can compare wave packets represented by Siegert pseudostates with confidence that discrepancy is due to 
error on the part of the Siegert pseudostates. We can do this using either Eq. (\ref{eq48a}) or Eq. (\ref{eq59}) with
Eq. (\ref{eq67}) to arrive at expansion coefficients.  For both forms we consider two cases: one where the initial wave 
packet has an average energy sufficient to pass the potential step at $r=a$ with little reflection ($k_0=15$), and one 
where $k_0=5$, ensuring significant physical reflection. The wave packets in the two cases are illustrated in 
Figs. \ref{fig3} and \ref{fig4}.  

\begin{table}
\caption[]{The time-dependent $\chi^2$'s discussed in Sec. \ref{sec3} for the case of the
fast-moving wave packet ($k_0=15$; see Fig. \ref{fig3}). $\chi_3^{2}(t)$ measures the quality of exponential time
evolution, the wave packet being expanded in a basis of bound and outgoing Siegert pseudostates
[Eq. (\ref{eq48a})]. $\chi_5^{2}(t)$ refers to nonexponential time evolution using all $2N$
Siegert pseudostates [Eqs. (\ref{eq59}), (\ref{eq67})]. $\parallel\psi(t)\parallel^2$ is defined in
Eq. (\ref{eq79}). The information in this table confirms that both expansions are accurate
in describing the motion of a higher-energy particle.}
\label{tab2}
\begin{ruledtabular}
\begin{tabular}{ccc}
$t$ [a.u.]  & $\chi_3^{2}(t)/\parallel\psi(t)\parallel^2$ & $\chi_5^{2}(t)/\parallel\psi(t)\parallel^2$ \\
\hline
$0$  & $6.2 \times 10^{-14}$ & $6.2 \times 10^{-14}$ \\
$0.125$ & $1.4 \times 10^{-13}$ & $1.4 \times 10^{-13}$  \\
$0.25$ & $2.2 \times 10^{-13}$ & $2.2 \times 10^{-13}$   \\
$0.375$ & $1.5 \times 10^{-12}$ & $1.5 \times 10^{-12}$  \\
$0.5$ & $1.9 \times 10^{-9}$ & $ 1.9 \times 10^{-9}$  \\
\end{tabular}
\end{ruledtabular}
\end{table}

\begin{table}
\caption[]{The case of the slow-moving wave packet ($k_0=5$; see Fig. \ref{fig4}).
$\chi_3^{2}(t)$ tests Eq. (\ref{eq48a}); $\chi_5^{2}(t)$ tests Eqs. (\ref{eq59}) and (\ref{eq67}).
$\parallel\psi(t)\parallel^2$ is defined in Eq. (\ref{eq79}). For both forms of expansion coefficients
[Eq. (\ref{eq48a}) vs. Eq. (\ref{eq67})], the accuracy at $t=0$ is only limited by the finite-element basis set.
For $t>0$, $\chi_5^2(t)$ is many orders of magnitude smaller than $\chi_3^2(t)$, showing that, for this case,
accurate time evolution is given by the expansion of Eq. (\ref{eq59}) using the nonexponential
time-evolution coefficients in Eq. (\ref{eq67}).}
\label{tab3}
\begin{ruledtabular}
\begin{tabular}{ccc}
$t$ [a.u.]  & $\chi_3^{2}(t)/\parallel\psi(t)\parallel^2$ & $\chi_5^{2}(t)/\parallel\psi(t)\parallel^2$ \\
\hline
$0$  & $2.9 \times 10^{-16}$ & $2.9 \times 10^{-16}$ \\
$0.5$ & $6.2 \times 10^{-6}$ & $5.3 \times 10^{-16}$  \\
$1$ & $5.5 \times 10^{-5}$ & $2.2 \times 10^{-15}$   \\
$1.5$ & $4.4 \times 10^{-4}$ & $2.4 \times 10^{-14}$  \\
$2$ & $ 1.2 \times 10^{-3}$ & $ 2.3 \times 10^{-13}$  \\
\end{tabular}
\end{ruledtabular}
\end{table}

In order to gauge the accuracy of Eq. (\ref{eq48a}) [i.e., exponential time evolution using a basis of $N$ Siegert
pseudostates], we modified $\chi_3^{2}$, Eq. (\ref{eq75}), to include
time dependence. We refer to this straightforward extension as $\chi_3^{2}(t)$. An analogous modification of
Eq. (\ref{eq78})---$\chi_5^{2}(t)$---was introduced to test the nonexponential time evolution expressed by 
Eqs. (\ref{eq59}) and (\ref{eq67}). Tables \ref{tab2} and 
\ref{tab3} were calculated with $N=620$. They compare time evolution using Eq. (\ref{eq48a}) with time evolution 
given by Eqs. (\ref{eq59}) and (\ref{eq67}) \cite{Faddee}. To account for decaying amplitude within the region 
$[0,a]$, we divide $\chi_3^{2}(t)$ and $\chi_5^{2}(t)$ by
\begin{equation}
\label{eq79}
\parallel\psi(t)\parallel^2 = \int_{0}^{a} \left|\psi(r;t)\right|^2 \mathrm{d}r \; .
\end{equation}

One sees in Table \ref{tab2} that for the case where $k_0 = 15$, Eq. (\ref{eq48a}) gives accurate time evolution, 
while Table \ref{tab3} shows a significant loss of precision for the slower wave packet, $k_0 = 5$, at all $t>0$.
This is because the assumption of pure exponential time evolution in Eq. (\ref{eq48a}) is not justified.
In fact, the more complicated time evolution suggested by Eq. (\ref{eq67}) provides superior agreement for both
$k_0 = 15$ and $k_0 = 5$. It is important to notice that the expansion coefficients associated with {\em all} Siegert 
pseudostates evolve in time in a {\em nonexponential} fashion. [Equation (\ref{eq48a}) fails also---for $k_0 = 5$---if 
$V_0$ is chosen to be so small that there are no bound states in the well.] Nevertheless, at least when $k_0=15$ (and 
bound and outgoing Siegert pseudostates are selected), Eq. (\ref{eq48a}) is satisfactory. In this case, we found 
numerically that the vector defined by its elements ${\bm c}_n^{\mathrm{T}} {\bm S} {\bm \alpha} = (\varphi_n|\psi)$, 
$n=1,\ldots,N$, is, to machine precision, an eigenvector of the matrix ${\bm M} \in \mathbb{C}^{N \times N}$
with eigenvalue one. As a consequence, the coefficients $\gamma_n$ in Eqs. (\ref{eq73}) and (\ref{eq75}) equal
$(\varphi_n|\psi)$. This means that Eq. (\ref{eq48a}) reduces to Eq. (\ref{eq80}).
We form the conjecture that this is the reason why the approach of Refs. \cite{YoWa99,TaWa01}, 
based on Eq. (\ref{eq80}), has been successful. [For $k_0 = 5$, we found the performance of Eq. (\ref{eq80})
to be even worse than that of Eq. (\ref{eq48a}).] The most general and accurate treatment, however, is only 
accomplished on the basis of the time evolution derived from the outgoing-wave Green's function, 
Eqs. (\ref{eq59}) and (\ref{eq67}).

\section{Conclusion}
\label{sec4}

In this paper, we reviewed the numerical technique developed by Tolstikhin, Ostrovsky, and Nakamura
\cite{ToOs97,ToOs98} for calculating Siegert pseudostates. The approach is applicable to general short-range 
potentials; it is not restricted to the simple model potential we made use of in our numerical investigation. 
We incorporated a straightforward extension of the method to nonorthogonal basis sets and explored in some detail
the mathematical properties of the Siegert pseudovectors, in particular questions of completeness. 

Our main goal in this paper has been the development of a rigorous formulation
of the time evolution of a wave packet expanded in terms of Siegert 
pseudostates. Our study demonstrates that assuming pure exponential time evolution can be a poor approximation.
The technique based on Eqs. (\ref{eq59}) and (\ref{eq67}), which implies nonexponential time evolution for the
entire set of $2N$ Siegert pseudostates, can reproduce very well the analytical reference wave packet moving in the
model potential. There are no artificial reflections at the boundary of the grid ($r=a$). Physical reflections
are correctly reproduced. 

We have shown that it is possible to find a subset consisting of $N$ Siegert pseudovectors that represents a basis of
$\mathbb{C}^N$. Siegert pseudostates were used in Ref. \cite{HaGr02} to carry out channel expansions in the context
of multichannel quantum defect theory, using bound and outgoing Siegert pseudostates only. This paper confirms
that such a choice provides a complete basis, in principle. In the time domain, this means that a wave packet can
be represented for any $t>0$ as a superposition of these $N$ basis states. Their time evolution is, however, 
nonexponential.
The time-dependent expansion coefficients may be found by applying the completeness relation implied by 
Eq. (\ref{eq47})  to the expansion in Eq. (\ref{eq59}).

While wave-packet propagation using Eqs. (\ref{eq59}) and (\ref{eq67}) is numerically stable and 
accurate, the calculation of {\em all} $2N$ eigenvalues and eigenvectors of the generalized eigenvalue problem
in Eq. (\ref{eq16}) is necessary in principle. This requirement poses a challenge for iterative sparse-matrix
techniques. Further developments will be needed in order to turn Siegert pseudostates into a useful tool for
large-scale calculations on atoms and molecules. Siegert pseudostates have already offered a glimpse at their 
extraordinary potential.

\acknowledgments
Financial support by the U.S. Department of Energy, Office of Science is gratefully acknowledged.

\appendix

\section{Existence of a subset of the $2N$ Siegert pseudovectors forming a basis of $\mathbb{C}^N$}
\label{appA}

Let $N_{\mathrm{C}}$ denote the number of eigenvectors ${\bm c}_n$ whose eigenvalue $\varkappa_n$ satisfies
$\mathrm{Im}(\varkappa_n) > 0$ [$\mathrm{Re}(k_n) > 0$]. There is an identical number of eigenvectors 
${\bm c}_n$ with $\mathrm{Im}(\varkappa_n) < 0$ [$\mathrm{Re}(k_n) < 0$]. The remaining $2N - 2N_{\mathrm{C}}$
eigenvectors are characterized by $\mathrm{Im}(\varkappa_n) = 0$ [$\mathrm{Re}(k_n) = 0$]. The latter correspond
to the bound and the antibound states. We order the $2N$ eigenvectors as follows:
\begin{equation}
\label{eq48}
\underbrace{{\bm c}_1,\ldots,{\bm c}_{N_{\mathrm{C}}}}_{\mathrm{Im}(\varkappa_n) > 0},
\underbrace{{\bm c}_{N_{\mathrm{C}}+1},\ldots,{\bm c}_{2N-N_{\mathrm{C}}}}_{\mathrm{Im}(\varkappa_n) = 0},
\underbrace{{\bm c}_{2N-N_{\mathrm{C}}+1},\ldots,{\bm c}_{2N}}_{\mathrm{Im}(\varkappa_n) < 0} \; .
\end{equation}
Additionally, the vectors ${\bm c}_{2N-N_{\mathrm{C}}+1},\ldots,{\bm c}_{2N}$ are ordered in such a way that
\begin{equation}
\label{eq49}
{\bm c}_{2N-N_{\mathrm{C}}+n} = {\bm c}_n^{\ast} \; , \; n = 1,\ldots,N_{\mathrm{C}} 
\end{equation}
[note the remarks following Eq. (\ref{eq12})]. We now describe a constructive approach for expressing the last 
$N_{\mathrm{C}}$ Siegert pseudovectors in Eq. (\ref{eq48}) [$\mathrm{Re}(k_n) < 0$] in terms of the vectors 
${\bm c}_1,\ldots,{\bm c}_{2N-N_{\mathrm{C}}}$.

Equations (\ref{eq35}) and (\ref{eq36}) can be multiplied from the right by ${\bm S}{\bm c}_m$. Let, in particular,
$1 \le m \le N_{\mathrm{C}}$. Then we can exploit that ${\bm c}_m^{\dag}{\bm S}{\bm c}_m > 0$ and solve for 
${\bm c}_m^{\ast}$:
\begin{eqnarray}
\label{eq50}
{\bm c}_m^{\ast} & = & 
-\sum_{n=1}^{2N-N_{\mathrm{C}}} \frac{\varkappa_m^{\ast}}{\varkappa_n}
\frac{{\bm c}_n^{\mathrm{T}}{\bm S}{\bm c}_m}{{\bm c}_m^{\dag}{\bm S}{\bm c}_m} {\bm c}_n \\
& & -\sum_{n=1 (n \ne m)}^{N_{\mathrm{C}}} \frac{\varkappa_m^{\ast}}{\varkappa_n^{\ast}}
\frac{{\bm c}_n^{\dag}{\bm S}{\bm c}_m}{{\bm c}_m^{\dag}{\bm S}{\bm c}_m} {\bm c}_n^{\ast} \; , \nonumber 
\end{eqnarray}
\begin{eqnarray}
\label{eq51}
{\bm c}_m^{\ast} & = & \frac{2}{{\bm c}_m^{\dag}{\bm S}{\bm c}_m} {\bm c}_m 
-\sum_{n=1}^{2N-N_{\mathrm{C}}} 
\frac{{\bm c}_n^{\mathrm{T}}{\bm S}{\bm c}_m}{{\bm c}_m^{\dag}{\bm S}{\bm c}_m} {\bm c}_n \nonumber \\
& & -\sum_{n=1 (n \ne m)}^{N_{\mathrm{C}}} 
\frac{{\bm c}_n^{\dag}{\bm S}{\bm c}_m}{{\bm c}_m^{\dag}{\bm S}{\bm c}_m} {\bm c}_n^{\ast} \; . 
\end{eqnarray}
According to these equations, ${\bm c}_{2N} = {\bm c}_{N_{\mathrm{C}}}^{\ast}$ can be represented in terms
of ${\bm c}_1,\ldots,{\bm c}_{2N-1}$, i.e.,
\begin{equation}
\label{eq52}
\mathrm{rank}([{\bm c}_1,\ldots,{\bm c}_{2N-1}]) = \mathrm{rank}([{\bm c}_1,\ldots,{\bm c}_{2N}]) = N \; .
\end{equation}

Let us assume now that, using Eqs. (\ref{eq50}) and (\ref{eq51}), we have been able to show for a fixed $m$,
$1 \le m \le N_{\mathrm{C}}$, that 
\begin{equation}
\label{eq53}
{\bm c}_{N_{\mathrm{C}}-n+1}^{\ast} = \sum_{k=1}^{2N-N_{\mathrm{C}}} \alpha_k^{(n)} {\bm c}_k
+ \sum_{k=1}^{N_{\mathrm{C}}-m} \beta_k^{(n)} {\bm c}_k^{\ast} \; , \; n=1,\ldots,m 
\end{equation}
for suitably chosen $\alpha_k^{(n)},\beta_k^{(n)}\in\mathbb{C}$. This is clearly true for $m=1$ [Eq. (\ref{eq52})].
It follows from Eq. (\ref{eq53}) that $\mathrm{rank}({\bm C}) = N$, where ${\bm C}=[{\bm c}_1,\ldots,{\bm c}_{2N-m}]$.
Also note that the coefficients $\alpha_k^{(n)},\beta_k^{(n)}$ cannot be unique (since $2N-m > N$), unless 
$m=N_{\mathrm{C}}$ and $N_{\mathrm{C}}=N$. In fact, there are an infinite number of solutions. This is easy to see:
Any solution ${\bm z}^{(n)}\in\mathbb{C}^{2N-m}$ of ${\bm C}{\bm z}^{(n)}={\bm c}_{N_{\mathrm{C}}-n+1}^{\ast}$
can be written as the sum of a particular solution of this linear system (which exists, by assumption) and a 
solution ${\bm y}\in\mathbb{C}^{2N-m}$ of the homogeneous system, ${\bm C}{\bm y} = {\bm 0}$.
The kernel of the matrix ${\bm C}$ is nontrivial, for its dimension is 
$\mathrm{dim}(\mathbb{C}^{2N-m})-\mathrm{rank}({\bm C}) = N-m$. Hence, there are infinitely many vectors ${\bm y}$
satisfying the homogeneous system.

We must make the step from $m$ to $m+1$. It is sufficient to demonstrate that
\begin{equation}
\label{eq54}
{\bm c}_{N_{\mathrm{C}}-m}^{\ast} = \sum_{k=1}^{2N-N_{\mathrm{C}}} \alpha_k^{(m+1)} {\bm c}_k
+ \sum_{k=1}^{N_{\mathrm{C}}-m-1} \beta_k^{(m+1)} {\bm c}_k^{\ast} \; .
\end{equation}
Equation (\ref{eq50}), applied to ${\bm c}_{N_{\mathrm{C}}-m}^{\ast}$, can be written as
\begin{eqnarray}
\label{eq55}
{\bm c}_{N_{\mathrm{C}}-m}^{\ast} & = & \sum_{n=1}^{2N-N_{\mathrm{C}}} (\ldots) {\bm c}_n
+ \sum_{n=1}^{N_{\mathrm{C}}-m-1} (\ldots) {\bm c}_n^{\ast} \\
& & -\sum_{n=N_{\mathrm{C}}-m+1}^{N_{\mathrm{C}}} \frac{\varkappa_{N_{\mathrm{C}}-m}^{\ast}}{\varkappa_n^{\ast}}
\frac{{\bm c}_n^{\dag}{\bm S}{\bm c}_{N_{\mathrm{C}}-m}}{{\bm c}_{N_{\mathrm{C}}-m}^{\dag}{\bm S}{\bm c}_{N_{\mathrm{C}}-m}} {\bm c}_n^{\ast} \; . \nonumber
\end{eqnarray}
Combined with Eq. (\ref{eq53}), this yields 
\begin{eqnarray}
{\bm c}_{N_{\mathrm{C}}-m}^{\ast} & = & \sum_{n=1}^{2N-N_{\mathrm{C}}} (\ldots) {\bm c}_n
+ \sum_{n=1}^{N_{\mathrm{C}}-m-1} (\ldots) {\bm c}_n^{\ast} \nonumber \\
& & -\sum_{n=N_{\mathrm{C}}-m+1}^{N_{\mathrm{C}}} \frac{\varkappa_{N_{\mathrm{C}}-m}^{\ast}}{\varkappa_n^{\ast}}
\frac{{\bm c}_n^{\dag}{\bm S}{\bm c}_{N_{\mathrm{C}}-m}}{{\bm c}_{N_{\mathrm{C}}-m}^{\dag}{\bm S}{\bm c}_{N_{\mathrm{C}}-m}} \nonumber \\
\label{eq56}
& & \times \beta_{N_{\mathrm{C}}-m}^{(N_{\mathrm{C}}-n+1)} {\bm c}_{N_{\mathrm{C}}-m}^{\ast}
 \; . 
\end{eqnarray}
The exact form of the coefficients symbolized by $(\ldots)$ is inessential. If we use Eq. (\ref{eq51}) in place
of Eq. (\ref{eq50}), we find
\begin{eqnarray}
{\bm c}_{N_{\mathrm{C}}-m}^{\ast} & = & \sum_{n=1}^{2N-N_{\mathrm{C}}} (\ldots) {\bm c}_n
+ \sum_{n=1}^{N_{\mathrm{C}}-m-1} (\ldots) {\bm c}_n^{\ast} \nonumber \\
& & -\sum_{n=N_{\mathrm{C}}-m+1}^{N_{\mathrm{C}}} 
\frac{{\bm c}_n^{\dag}{\bm S}{\bm c}_{N_{\mathrm{C}}-m}}{{\bm c}_{N_{\mathrm{C}}-m}^{\dag}{\bm S}{\bm c}_{N_{\mathrm{C}}-m}\
} \nonumber \\
\label{eq57}
& & \times \beta_{N_{\mathrm{C}}-m}^{(N_{\mathrm{C}}-n+1)} {\bm c}_{N_{\mathrm{C}}-m}^{\ast} \; . 
\end{eqnarray}
Therefore, as long as either 
\[
-\sum_{n=N_{\mathrm{C}}-m+1}^{N_{\mathrm{C}}} \frac{\varkappa_{N_{\mathrm{C}}-m}^{\ast}}{\varkappa_n^{\ast}}
\frac{{\bm c}_n^{\dag}{\bm S}{\bm c}_{N_{\mathrm{C}}-m}}{{\bm c}_{N_{\mathrm{C}}-m}^{\dag}{\bm S}{\bm c}_{N_{\mathrm{C}}-m}} \beta_{N_{\mathrm{C}}-m}^{(N_{\mathrm{C}}-n+1)}
\]
or
\[
-\sum_{n=N_{\mathrm{C}}-m+1}^{N_{\mathrm{C}}}
\frac{{\bm c}_n^{\dag}{\bm S}{\bm c}_{N_{\mathrm{C}}-m}}{{\bm c}_{N_{\mathrm{C}}-m}^{\dag}{\bm S}{\bm c}_{N_{\mathrm{C}}-m}} \beta_{N_{\mathrm{C}}-m}^{(N_{\mathrm{C}}-n+1)}
\]
differ from one, Eq. (\ref{eq54}) is satisfied. Otherwise, since the coefficients $\beta_k^{(n)}$ are not unique,
it is possible in general to choose the $\beta_{N_{\mathrm{C}}-m}^{(N_{\mathrm{C}}-n+1)}$ such that either
Eq. (\ref{eq56}) or (\ref{eq57}) can be solved for ${\bm c}_{N_{\mathrm{C}}-m}^{\ast}$.

The coefficients $\beta_{N_{\mathrm{C}}-m}^{(N_{\mathrm{C}}-n+1)}$ will be unique, however, if and only if
\begin{equation}
\label{eq58}
\mathrm{rank}([{\bm c}_1,\ldots,{\bm c}_{2N-m}]) = \mathrm{rank}([{\bm c}_1,\ldots,{\bm c}_{2N-m-1}]) + 1 \; ,
\end{equation}
which would also be consistent with the disappearance from Eqs. (\ref{eq56}) and (\ref{eq57}) of the term
involving ${\bm c}_{N_{\mathrm{C}}-m}^{\ast}$. Thus, in exact arithmetic, it is conceivable that the construction step
from $m$ to $m+1$ fails, i.e., we cannot prove that {\em all} incoming Siegert pseudovectors can be eliminated.
(Nevertheless, there is no reason to anticipate this to cause any difficulties in numerical
calculations.) If really necessary, then the vector ${\bm c}_{N_{\mathrm{C}}-m}^{\ast}$ must be kept as an essential
basis vector, and the construction procedure outlined above may be continued for the remaining vectors
${\bm c}_1,\ldots,{\bm c}_{2N-m-1}$.

Not only the vectors ${\bm c}_n^{\ast}$ [$\mathrm{Re}(k_n) < 0$] can be eliminated (except for the possible---though
unlikely---occurrence of essential vectors), but equations analogous to Eqs. (\ref{eq50}) and (\ref{eq51}) exist 
also for the bound and the antibound eigenvectors. Note that for these, ${\bm c}_m^{\mathrm{T}}{\bm S}{\bm c}_m$ 
differs from $0$. It is therefore clear that $N$ linearly independent basis vectors ${\bm c}_n$, $n=1,\ldots,N$, 
can be found among the vectors ${\bm c}_1,\ldots,{\bm c}_{2N}$. The completeness relation satisfied by these $N$ 
vectors is given by Eq. (\ref{eq47}).

\section{Time evolution and the Faddeeva function}
\label{appB}

For evaluating the integral in Eq. (\ref{eq64}), it is useful to perform the substitution 
\begin{equation}
\label{eqB1}
s^2 = \mathrm{i} k^2 t/2 \; .
\end{equation}
Thus,
\begin{equation}
\label{eqB2}
\int_{-\infty}^{\infty} \frac{\mathrm{e}^{-\mathrm{i} k^2 t/2}}{k-k_n} \mathrm{d}k
=
\int_{-\sqrt{\mathrm{i}t/2}\infty}^{+\sqrt{\mathrm{i}t/2}\infty} 
\frac{\mathrm{e}^{-s^2}}{s-\sqrt{\mathrm{i}t/2}k_n} \mathrm{d}s \; .
\end{equation}
The right-hand side of this equation is reminiscent of the Faddeeva function defined in Eq. (\ref{eq66}).
However, the integration path must be rotated back to the real axis. In order to do this, we must
carefully distinguish between poles due to bound, antibound, outgoing, or incoming Siegert pseudostates.
The pole of the integrand on the right-hand side of Eq. (\ref{eqB2}) lies in the upper $s$ plane if, 
for $t>0$ ($\sqrt{t/2}>0$),
\begin{equation}
\label{eqB3}
\sqrt{\frac{\mathrm{i}t}{2}} = \mathrm{e}^{\mathrm{i} \pi / 4}\sqrt{\frac{t}{2}}
\end{equation}
for bound or outgoing Siegert pseudostates, and
\begin{equation}
\label{eqB4}
\sqrt{\frac{\mathrm{i}t}{2}} = -\mathrm{e}^{\mathrm{i} \pi / 4}\sqrt{\frac{t}{2}}
\end{equation}
for antibound or incoming Siegert pseudostates. (Note that the Faddeeva function is defined in the upper
half-plane.) 

\begin{figure}
\includegraphics[width=4.5cm,origin=c,angle=0]{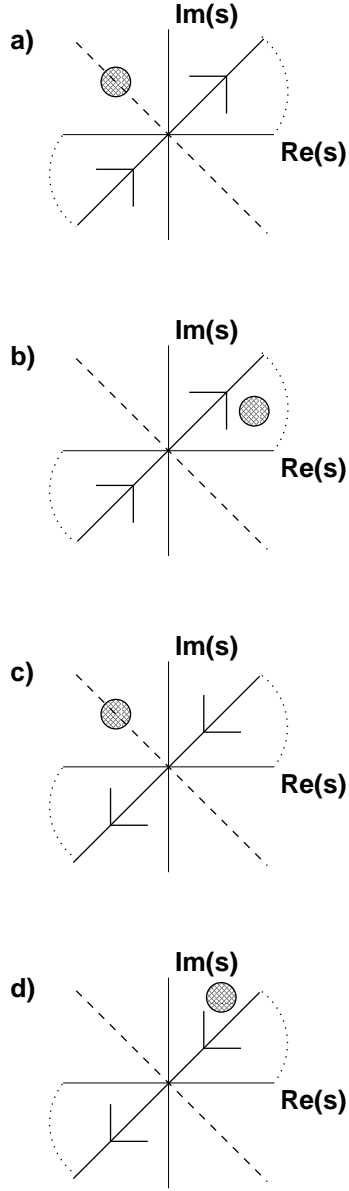}
\caption[]{The integration contour used to evaluate Eq. (\ref{eqB2}) depends on the nature of the Siegert pseudostate
considered. Arrows indicate the direction of the integration path in the complex $s$ plane.  The circle symbolizes
the location of the pole of the integrand on the right-hand side of Eq. (\ref{eqB2}).
a) Bound state. b) Outgoing Siegert pseudostate. c) Antibound state. d) Incoming Siegert pseudostate.}
\label{figB1}
\end{figure}

Let us first consider a bound state. In this case, the pole is in the second quadrant of the complex
$s$ plane. The integration path, illustrated in Fig. \ref{figB1}a together with the location of the bound-state pole,
runs along the diagonal from the third to the first quadrant. Since the contour may be closed as indicated in the
figure, and since there are no poles inside the resulting loop, we find
\begin{eqnarray}
\int_{-\sqrt{\mathrm{i}t/2}\infty}^{+\sqrt{\mathrm{i}t/2}\infty}
\frac{\mathrm{e}^{-s^2}}{s-\sqrt{\mathrm{i}t/2}k_n} \mathrm{d}s 
& = & \int_{-\infty}^{\infty} \frac{\mathrm{e}^{-s^2}}{s-\mathrm{e}^{\mathrm{i} \pi / 4}\sqrt{t/2}k_n} \mathrm{d}s 
\nonumber \\ 
\label{eqB6}
& = & -\frac{\pi}{\mathrm{i}} w\left(\mathrm{e}^{\mathrm{i} \pi / 4}\sqrt{\frac{t}{2}}k_n\right)
\end{eqnarray}
for a bound state.

The location of the pole in the case of an outgoing Siegert pseudostate is shown in Fig. \ref{figB1}b. This time the
pole lies inside the integration loop, so that
\begin{eqnarray}
\label{eqB7}
\int_{-\sqrt{\mathrm{i}t/2}\infty}^{+\sqrt{\mathrm{i}t/2}\infty}
\frac{\mathrm{e}^{-s^2}}{s-\sqrt{\mathrm{i}t/2}k_n} \mathrm{d}s
& = & -2\pi\mathrm{i} \mathrm{e}^{-\mathrm{i} k_n^2 t/2} \\
&   & -\frac{\pi}{\mathrm{i}} w\left(\mathrm{e}^{\mathrm{i} \pi / 4}\sqrt{\frac{t}{2}}k_n\right) \nonumber
\end{eqnarray}
for an outgoing Siegert pseudostate.

The pole's location for an antibound and an incoming Siegert pseudostate, respectively, can be seen in Figs. \ref{figB1}c
and \ref{figB1}d. (The integration path now runs along the diagonal from the first to the third quadrant.)
In neither case does the pole interfere with the rotation of the integration path to the real $s$ axis. Therefore,
for an antibound or incoming Siegert pseudostate the result reads
\begin{eqnarray}
\int_{-\sqrt{\mathrm{i}t/2}\infty}^{+\sqrt{\mathrm{i}t/2}\infty}
\frac{\mathrm{e}^{-s^2}}{s-\sqrt{\mathrm{i}t/2}k_n} \mathrm{d}s
& = & -\int_{-\infty}^{\infty} \frac{\mathrm{e}^{-s^2}}{s+\mathrm{e}^{\mathrm{i} \pi / 4}\sqrt{t/2}k_n} \mathrm{d}s 
\nonumber \\
\label{eqB9}
& = & \frac{\pi}{\mathrm{i}} w\left(-\mathrm{e}^{\mathrm{i} \pi / 4}\sqrt{\frac{t}{2}}k_n\right) \; .
\end{eqnarray}


\begin{thebibliography}{99}
\bibitem{Faddee} The Faddeeva function in Eq. (\ref{eq67}) was numerically evaluated using the
algorithm described in Ref. \cite{PoWi90}.
\end{thebibliography}
\end{document}